\documentclass{iau} 
\usepackage{graphicx}
 
\title[Super star cluster driven feedback] 
{The super star cluster driven feedback in ESO338-IG04 and Haro 11}

\author[A. Bik  et al.]   
{A. Bik$^1$,
 G. {\"O}stlin$^1$,
 V.  Menacho$^1$,
 A. Adamo$^1$,
 M. Hayes$^1$,
 J. Melinder$^1$,
\and P. Amram$^2$}

\affiliation{$^1$Department of Astronomy, Oskar Klein Centre, Stockholm University, \\AlbaNova University Centre, 106 91 Stockholm, Sweden \\ email: {\tt arjan.bik@astro.su.se} \\[\affilskip]
$^2$Aix Marseille Universit\'{e}, CNRS, LAM (Laboratoire dÕAstrophysique de Marseille), 13388 Marseille, France}

\pubyear{2015}
\volume{316}  
\jname{Title of your IAU Symposium}
\editors{A.C. Editor, B.D. Editor \& C.E. Editor, eds.}
\begin{document}

\maketitle

\begin{abstract}

The stellar content of young massive star clusters emit large amounts of Lyman continuum photons and inject momentum into the inter stellar medium (ISM) by the strong stellar winds of the most massive stars in the cluster. When the most massive stars explode as supernovae, large amounts of mechanical energy are injected in the ISM. A detailed study of the ISM around these massive cluster provides  insights on the effect of cluster feedback.

We present high quality integral field spectroscopy taken with VLT/MUSE of two starburst galaxies: ESO 338-IG04 and Haro 11. Both galaxies contain a significant number of super star clusters. The MUSE data provide us with an unprecedented view of the state and kinematics of the ionized gas in the galaxy allowing us to study the effect of stellar feedback on small and large spatial scales. We present our recent results on studying the ISM state of these two galaxies. The data of both galaxies show that the mechanical and ionization feedback of the super star clusters in the galaxy modify the state and kinematics of the ISM substancially by creating highly ionized bubbles around the cluster, making the central part of the galaxy highly ionized. This shows that the HII regions around the individual clusters are density bounded, allowing the ionizing photons to escape and ionize the ISM further out.


\end{abstract}
 
\firstsection 
\section{Introduction}

Young massive stars and  star clusters have a strong impact on the gas in their host galaxy. Young star star clusters, containing many massive stars, emit large amounts of Lyman continuum photons (LyC) able to ionize the ISM.  Additionally, momentum is injected by the strong stellar winds of the most massive stars in the cluster. For slightly older clusters, (3 Myrs or older), the most massive stars explode as supernovae enabling the injection of large amounts of mechanical energy into the ISM. 

These different feedback mechanisms have different effects on the ISM surrounding the clusters. Optical integral field spectroscopy with MUSE provides the ideal tool for studying affect the clusters have on the ISM in the galaxy. The youngest clusters will ionize their surroundings, resulting in a highly ionized medium around the clusters, traced by the ionization parameter (\cite[e.g. Pellegrini \etal\ 2012]{Pellegrini12}). The mechanical feedback from stellar winds and supernovae can result in galaxy scale outflows, traced by the kinematics of e.g. the H$\alpha$ line.

\begin{figure}[t]
\begin{center}
\includegraphics[width=13.2cm]{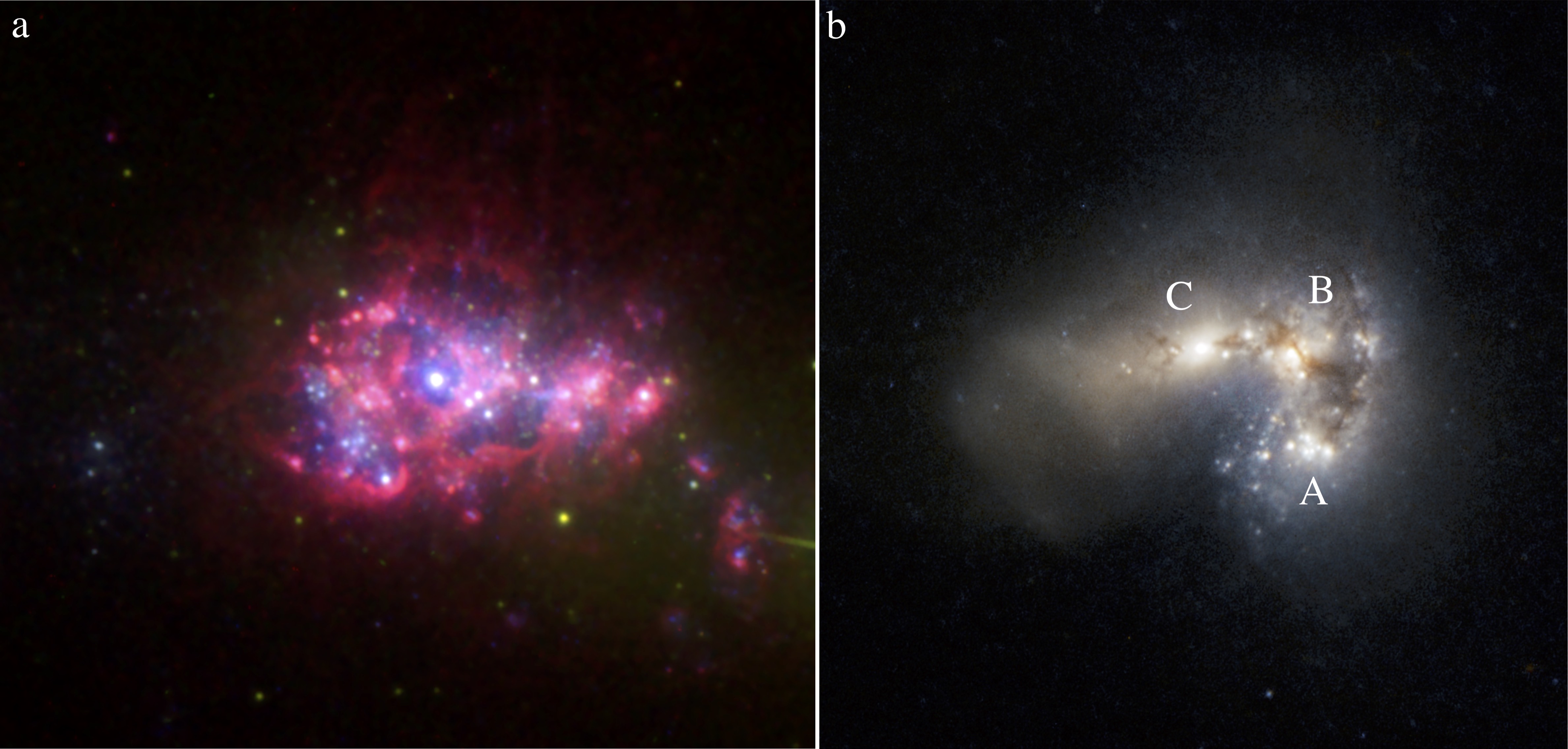} 
 \caption{{\bf a)} Three color-composite of HST observations of ESO 338-IG04. Blue: F140LP (UV), Green: F550M (V-band), Red: FR656N (H$\alpha$) The field of view of this image of 20 $\times$ 18 arcsec. {\bf b)} Three color-composite HST observations of Haro 11, Blue: B-band, Green: I-band, Red: K-band. (Taken from \cite{Adamo10}. Credit: ESO/ESA/Hubble and NASA). The field of view of this image of 43 $\times$ 43 arcsec.}
   \label{fig:HST}
\end{center}
\end{figure}

We present optical integral field spectroscopy of two blue compact galaxies, Haro 11 and ESO 338-IG04 (Fig \ref{fig:HST}). These galaxies are considered local analogues of the high redshift Lyman break galaxies (\cite{Overzier08}). Studying their star formation and ISM properties teaches us about the star formation process in the  high redshift universe.

Both galaxies have  well studied stellar cluster populations, containing tens of super star clusters (\cite{Ostlin98,Ostlin03,Adamo10,Adamo11}). 
ESO 338 contains a very young stellar population where the current starburst started $\sim$ 40 years ago (\cite{Ostlin03}). Several increases in the cluster formation history have been found, the earliest at $\sim$ 20 - 30 Myrs, while the latest increase in cluster formation rate started 10 Myrs ago (\cite{Adamo11}). In Haro 11 the star formation is concentrated in 3 knots (\cite{Vader93}), each containing tens of young and older super star clusters. These galaxies show a high cluster formation efficiency, suggesting that 50 \% of the star formation happens in the form of bound stellar clusters (\cite{Adamo11}). These properties make the galaxies ideal targets to study the effect of super star cluster feedback on the ISM of the galaxy.


\section{Observations}

\begin{figure}[t]
\begin{center}
\includegraphics[width=13.2cm]{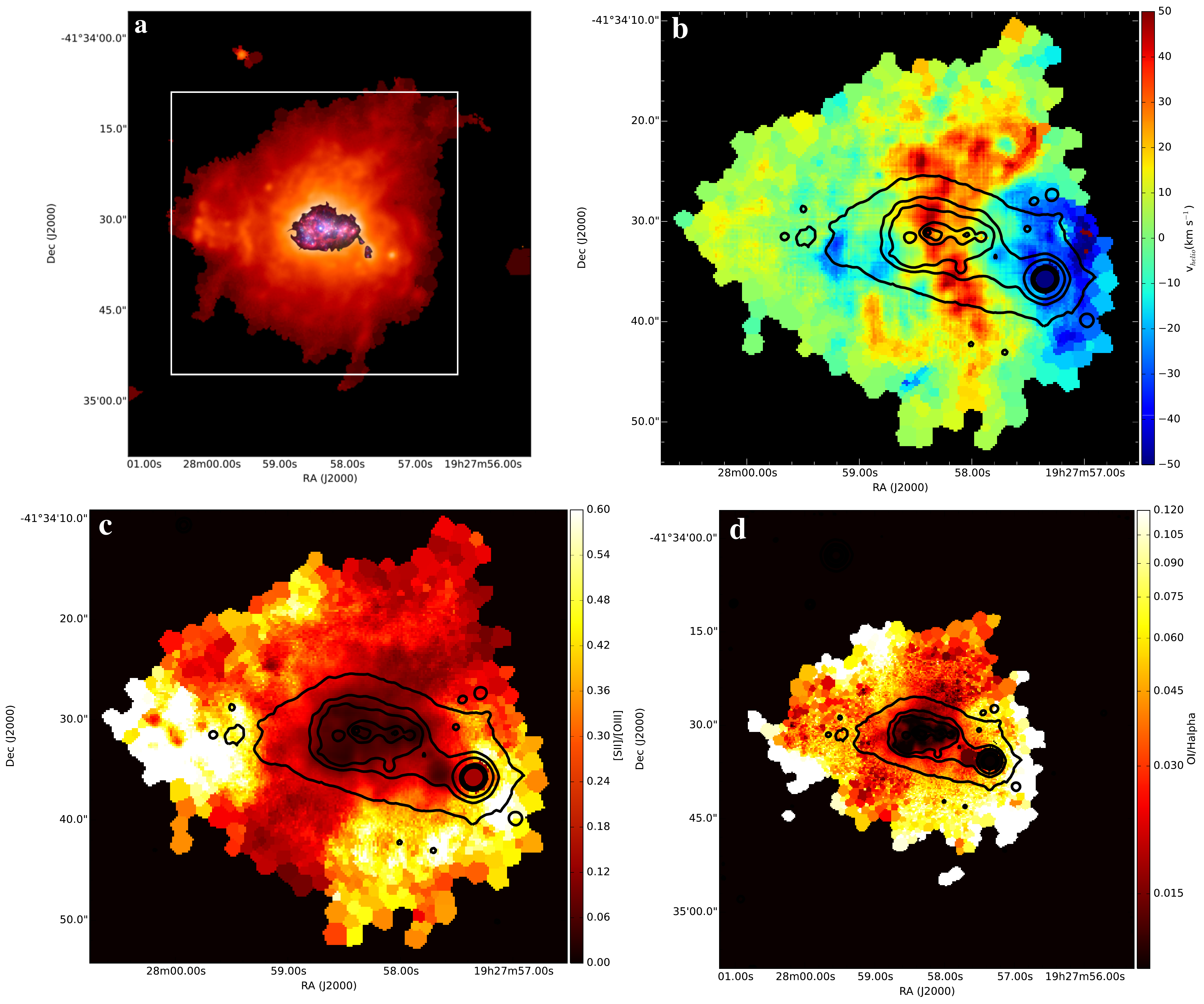} 
 \caption{MUSE observations of ESO338-IG04. a) Continuum subtracted H$\alpha$ with the HST image of Fig \ref{fig:HST} super imposed, demonstrating the large halo surrounding ESO 338-IG04. The white box shows the zoomed in portion of the data visible in the other panels. b) The H$\alpha$ velocity field, indicting large scale redshifted outflows are present. The black contours are contours of continuum emission, derived from stacking the MUSE cube along the wavelength direction. c) The ionization parameter represented by the [SII]/[OIII] line ratios showing the presence of ionization cones. d) The OI/H$\alpha$ ratio indicting the presence of shocks at the other edges of the halo.}
   \label{fig:E338}
\end{center}
\end{figure}

\begin{figure}[t]
\begin{center}
\includegraphics[width=13.2cm]{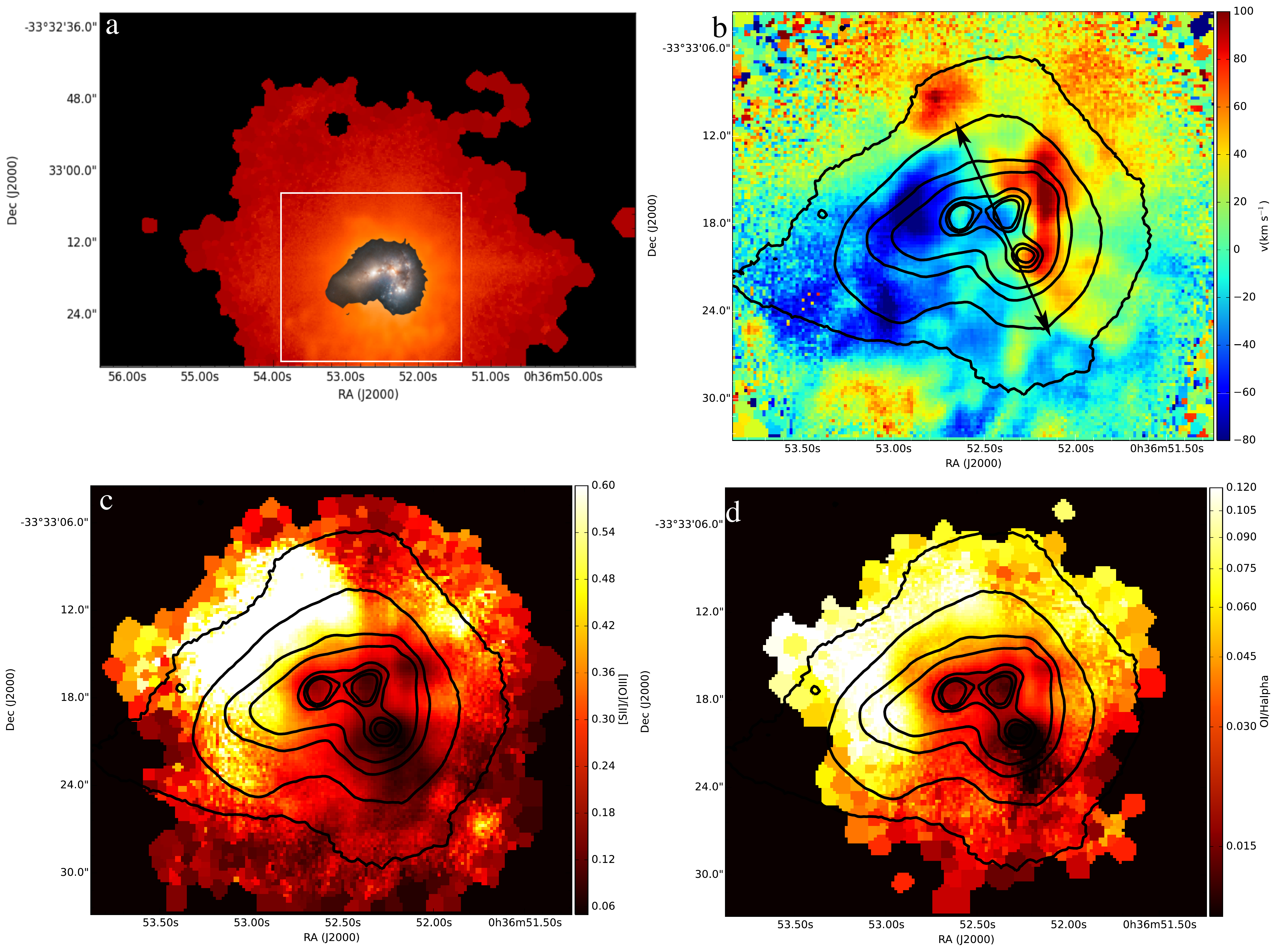} 
 \caption{MUSE observations of Haro 11, panels present the same information as in Fig. 1. The arrow in panel b highlights the 2 suspected outflows in Haro 11.}
   \label{fig:H11}
\end{center}
\end{figure}

 ESO 338 and Haro 11 were observed with the integral field spectrograph Multi-Unit Spectroscopic Explorer (MUSE; \cite[Bacon \etal\ 2010]{Bacon10}) mounted on the Very Large Telescope (VLT) on Paranal, Chile. The observations of ESO 338  were performed during the first science verification run (\cite{Bik15}), the observations of Haro 11 as part of regular service mode observations. 
 
The observations are taken in the no-AO mode in the extended wavelength setting, providing spectra between 4650 and 9350 \AA\ in a field of view of 1 x 1 arcmin with a 0.2 arcsec pixel scale. One pointing, consisting of 4 integrations of 750 seconds, each rotated with 90 degrees, was used to observe ESO 338-IG04. The typical image quality is 0.9 arcsec in the V-band. For Haro 11 two pointings, offset with 30 arcsec east-west were obtained, to cover the large extent of the ionized halo. The image quality here is 0.8 arcsec.  The data were reduced using the ESO MUSE pipeline (Weilbacher et al, in prep). 

Emission line maps were extracted from the final data by numerically integrating under the line profile. The continuum was subtracted by averaging the continuum red- and blueward of the line. To enhance the low-surface brightness features, we binned the data using the weighted Voronoi tesselation algoritm by \cite{Diehl06}, which is a generalization of the algoritm developed by \cite{Cappellari03}.  Additionally, the velocity field of the H$\alpha$ line was calculated by fitting a single  Gaussian profile.

\section{Results}

Figs.\,\ref{fig:E338} and \ref{fig:H11} show the results obtained from the MUSE data. Panels (a) show the H$\alpha$ intensity maps and demonstrate that both galaxies are surrounded by large ionized halos (14.5 kpc for Haro 11 and 4.5 kpc for ESO 338). The velocity map (Figs.\,\ref{fig:E338}b  and \ref{fig:H11}b) shows that the velocity field of the ionized gas is not simply tracing the gravitational potential of the galaxy. ESO 338 shows evidence for several redshifted outflows (\cite{Bik15}). The velocity map of Haro 11  looks very complicated and most of the gas dynamics is related to the merger event (\cite{Ostlin15}). A detailed comparison between the H$\alpha$ intensity map and velocity map reveals a blue  and redshifted outflow candidate (annotated in Fig  \ref{fig:H11}b). 

To study the level of ionization in the ISM we calculated the [SII]($\lambda$6717 \AA + $\lambda$6731 \AA)/[OIII]$\lambda$5007\AA\ line ratio (Fig.\,\ref{fig:E338}c  and \ref{fig:H11}c). The value of this line ratio is related to the ionization parameter and has been calibrated by \cite{Pelligrini12}. The lowest values of the line ratios are found towards the location of the young star clusters, consistent with the presence of early O stars in these clusters. However, also on larger scale, a significant fraction of the halos is ionized. In ESO 338 two ionization cones are cleary visible (to the northwest and south east). The ionization parameter map of Haro 11 shows that the Halo is ionized towards the west and south of the galaxy. Only towards the north-east is ISM the neutral.  This means that the halos of both galaxies are density bound and facilitate the leakage of LyC photons.

In  Fig.\, \ref{fig:E338}d  and \ref{fig:H11}d the OI$\lambda$6300\AA/H$\alpha$ line ratio is shown, allowing us to discriminate whether the gas is heated by UV photo ionization (low values) or by shocks (high values). As expected from the ionization parameter maps, the location of the clusters show very low values, suggesting the heating by UV photo ionization of the young massive stars. However, where the material becomes more neutral the OI/H$\alpha$ line ratio suggests the presence of shocks. In ESO 338, the high values are observed towards the outer edges of the halo, while in Haro 11 the north-east side of the halo is strongly shocked.

\begin{figure}[t]
\begin{center}
\includegraphics[width=13.2cm]{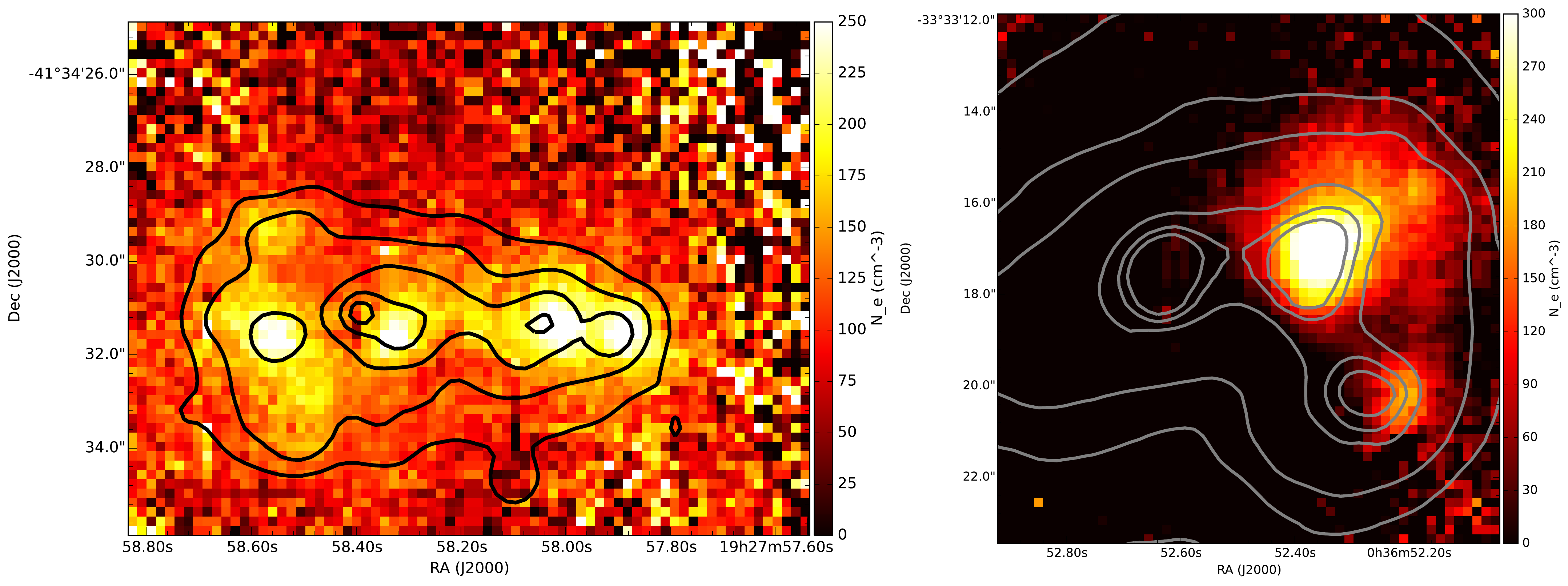} 
 \caption{Electron density (n$_{e}$) derived from the line ratio maps of [SII]$\lambda$6717\AA\ and [SII]$\lambda$6731\AA. The  contours are the same as in Fig. \ref{fig:E338}\emph{Left:} Electron density map of the inner regions of the halo of ESO 338. The HII regions surrounding the young clusters reach higher densities then the low-density halo gas. \emph{Left:}  The electron density map of Haro 11. Knot B shows up most prominent as it contains most of the young cluster (\cite{Adamo10}).}
   \label{fig:Ne}
\end{center}
\end{figure}

From the  line ratio [SII]$\lambda$6717\AA/[SII]$\lambda$6731\AA\ we calculate the electron density in the halos and the HII regions surrounding the clusters using the calibration of \cite{Osterbrock98}. Fig \ref{fig:Ne} shows that the halo gas around the galaxies is of low density (n$_{e} <$ 100 cm$^{-3}$)  where the observed line ratio reaches the asymtotic value of the calibration and cannot be used to derive quantative densities. Towards the HII regions surrounding the clusters in the galaxies, however, we find electron densities of n$_{e}$ =  200 - 400 cm$^{-3}$. This suggests that these young clusters are still surrounded by parts of their natal HII region.

\section{Discussion}

\subsection{  LyC and Ly$\alpha$ escape}

The ionization parameter maps show that the halos of both galaxies are density bound over a large part of their halos. The highly ionized cones of ESO 338 as well as the southern and western part of Haro 11 facilitate the escape of LyC photons in the intergalactic medium (see also \cite{Zastrow11} and \cite{Zastrow13}). In fact Haro 11 is one of the few nearby galaxies where the LyC continuum radiation has been detected directly (\cite{Leitet11}).  For ESO 338 an estimate of a LyC escape fraction of  16 \% was derived based on analysis of the CII $\lambda$1036\AA\ absorption line (\cite{Leitet13}).

Comparing the ionization cones and the outflows detected in ESO 338 with the Ly$\alpha$ emission maps of \cite{Hayes07}, a spatial correlation was found between the asymmetries  in the Ly$\alpha$ map and the location of the outflows, suggesting that the outflows facilitate the escape of Ly$\alpha$ emission (\cite{Bik15}).

\subsection{stellar cluster feedback}

The observed properties of the ISM as discussed above can be linked to the star cluster population of ESO 338 and Haro 11. The ionization maps of both galaxies show several small regions of highly ionized gas. Matching their locations with those of the young stellar clusters shows that in the case of ESO 338 these small regions are highly ionized HII regions with high electron density, surrounding the youngest clusters. \cite[\"Ostlin et al, (2003)]{Ostlin03} derive ages of less than 2 Myrs for those clusters.  The data on Haro 11 show a similar picture, all the star forming knots are surrounded by highly ionized, high density HII regions, consistent with them containing young star clusters (\cite{Adamo10}), with one exception, however. Knot C is detected in the ionization map, but shows a low electron density, like the rest of the halo around Haro 11. This is consistent with the fact that knot C is the older of the star forming knots, and the ionization might be caused by something other than star formation.

On larger spatial scales, the data show a low-density, partly highly ionized halo around both galaxies. ESO 338 shows clear presence of outflows as well as two ionization cones. Comparing the alignment of the outflows with that of the ionization cones shows that the southern outflow is not aligned with the southern ionization cone, and are probably two unrelated features.  LyC photons, capable of ionizing the ISM are only produced by clusters with ages younger than $\sim 4$ Myrs, while the mechanical energy released by stellar winds and supernovae remains roughly constant for $\sim 30$ Myrs (\cite{Leitherer99}). This means that clusters with ages of $\sim 30$ Myrs have inject most mechanical energy in the ISM and therefore might be responsible for the observed outflows. The LyC photons responsible for the ionized cones are created by the most recent increase in star formation. These channels might be outflow cones created by the older generation of clusters and might be re-ionized by the younger generation.

\section{Conclusion}

Both galaxies show that the ISM is significantly altered by the feedback from the super star clusters. Apart from the small HII regions around each young clusters, the ionization  makes the galaxy act like a giant HII region expanding into the more neutral outer regions of ISM and creating shocks visible in the OI/H$\alpha$ line ratio. These results confirm the results of \cite{Cormier12} for Haro 11 derived from mid- and  far-infrared fine structure lines. They built up a multi-phase ISM model of Haro 11, consisting of dense, highly ionized HII regions around the individial star clusters. These are surrounded by fragmented photo dissociation regions and embedded in a low-density, low ionized halo. With the MUSE data we confirm this picture and provide the missing two dimensional information on the location of the different ISM phases in the galaxies.


\end{document}